\documentclass[aps,prl,twocolumn,showpacs]{revtex4}

\usepackage{amsmath,amsfonts,amssymb}
\usepackage{psfrag}
\usepackage{graphicx}

\newcommand{\nop}[1]{}

\begin{document}


\title{A logarithmic contribution to the density of states 
of rectangular Andreev billiards}

\author{A. Korm\'anyos}
\email{kor@complex.elte.hu}
\author{Z. Kaufmann}
\author{J. Cserti}
\affiliation{Department of Physics of Complex Systems, 
E{\"o}tv{\"o}s University
\\ H-1117 Budapest, P\'azm\'any P{\'e}ter s{\'e}t\'any 1/A, Hungary}
\author{C. J. Lambert}
\affiliation{Department of Physics, Lancaster University,
Lancaster, LA1 4YB, UK}


\begin{abstract}
 
We demonstrate that the exact quantum mechanical calculations are in  
good agreement with the semiclassical predictions for rectangular 
Andreev billiards and therefore for a large number of open channels it is
sufficient to investigate the Bohr-Sommerfeld approximation of 
the density of states. 
We present exact calculations of the classical  path length
distribution $P(s)$ which is a non-differentiable function of $s$,
but whose integral is a smooth function with logarithmically dependent
asymptotic behavior. Consequently, the 
density of states of rectangular Andreev billiards has two contributions 
on the scale of the Thouless energy: one which is  well-known and 
it is proportional to the energy, and the other which shows a logarithmic 
energy dependence.  It is shown that the prefactors of both 
contributions depend on the geometry of the billiards but they have 
universal limiting values when the width of the superconductor tends to zero.

\end{abstract}


\maketitle


When a normal-metallic dot is placed in contact with a superconductor, 
the low-energy density of states $n(E)$ of the resulting Andreev billiard 
is strongly modified compared with the normal state \cite{Andreev-Billiards}. 
The energy dependence
of $n(E)$ is highly non-trivial and provides a testing ground for current
understanding of proximity effects of hybrid superconducting nanostructures 
and for theoretical tools such as semiclassical theory.
Recently a number of conflicting results have been obtained for the density of 
states (DOS) of ballistic Andreev billiards (AB). One example is shown 
in figure 1, 
which in the separable limit $W=a$ was studied long ago 
in \cite{deGennes-Saint-James}, where
it was found that
for small $E$, $n(E)$ is proportional to $E$. In contrast for $W<a$,
 the problem has only recently been 
studied \cite{Andreev-Billiards,Melsen1,Richter1,box_disk:cikk,Gap-cikk}. 
In Ref.~\cite{Melsen1} it was found that $n(E)=\nu E$, where $\nu$ is a
universal 
constant, independent 
of $E$ and $W$. In contrast \cite{Richter1} predicted different values 
for $\nu$. 

In this paper we resolve this discrepancy by showing that neither 
of these results is strictly correct in the limit $E\rightarrow 0$ and 
instead predict that $n(E)$ diverges logarithmically. This generic 
logarithmic contribution
is significant for energies less than or of order the Thouless energy.

To address this problem, we use both exact and semiclassical techniques to
analyze the AB of figure 1. 
The exact calculation starts with a ballistic two dimensional normal
dot of area $A$, described by a scattering matrix $S_0(E)$ and with
a mean level spacing for the isolated normal system $\delta
=\frac{2\pi \hbar^2}{mA}$ at the Fermi energy $E_F$. 
Then, a  superconductor of width $W$ and bulk order parameter $\Delta$ is
placed in contact with such a billiard. 
The number of open channels in the S region is the integer part of
$M=\frac{k_{\rm F}W}{\pi}$, and the energy levels of AB 
are the positive eigenvalues $E$ (measured from the Fermi energy) 
of the Bogoliubov-de Gennes equation\cite{BdG-eq}.
A secular equation of the {\em exact} energy levels of 
AB, in terms of the scattering matrix $S_0(E)$ of the normal region, is 
derived by matching the wave functions at the interface of the
normal-superconductor (N-S) systems. The energy levels are exact 
in the sense that no Andreev approximation ($\Delta/E_{\rm F} \ll 1$
and quasi-particles whose incident/reflected directions are approximately
perpendicular to the N-S interface)~\cite{Colin-review} is assumed. 

\begin{figure}[hbt]
\includegraphics[scale=0.3]{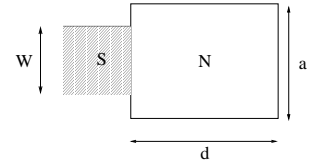}
\caption{A normal dot of rectangular shape in contact with a
superconductor placed at the middle of the edge of the normal region. 
\label{geometria-fig}}
\end{figure}

We also give the semiclassical Bohr-Sommerfeld 
approximation of the density of states (DOS) $n(E)$ expressed by 
the classical return probability $P(s)$ of the electron. 
Our exact and semiclassical calculations are applied to the  
{\em integrable}  AB shown in Fig.~\ref{geometria-fig}. 
We demonstrate that the integrated DOS agrees very well with 
the exact calculations. 
We find that the integral of the return probability $P(s)$ has a 
contribution depending on $s$ {\em logarithmically} 
in the asymptotic limit, $s \to \infty$, which has to-date been overlooked.
 As a consequence, 
the small energy dependence of the DOS  $n(E)$ has also a 
{\em logarithmic factor} in addition to a contribution which  
depends linearly on the energy $E$ as predicted 
in Ref.~\onlinecite{Melsen1,Richter1}. 

In our earlier work~\cite{Gap-cikk} we found that the exact energy levels 
of AB with arbitrary shape of normal region 
are the solutions of the following secular equation: 
\begin{subequations}
\label{exact-QM:e}
\begin{equation}
\det \Bigl(\rm Im \bigl\{  \gamma D_e(E) D_h(E) \bigr \} \Bigr) 
= 0, \,\, where 
\label{seceq:e}
\end{equation} 
\begin{eqnarray}
D_e(E) &=&  Q(E)+K(E)G(E),    \\
D_h(E) &=& {\bigl[Q(-E)-K(-E)G^*(-E)\bigr]}^{-1},  \\
G(E) &=& \left[1-S_0(E)\right]{\left[1+S_0(E)\right]}^{-1}.   \label{G_def:e}
\end{eqnarray}
\end{subequations}  
Here $\gamma=e^{-i\arccos \, \left(E/\Delta\right) }$, 
$Q$ and $K$ are diagonal matrices with elements
$Q_{nm}(E) = \delta_{nm}\,q_n(E)$ and 
$K_{nm}(E) = \delta_{nm}\, k_n(E)$, where 
$q_n (E)=k_F \sqrt{1 + i\, 
\frac{\sqrt{\Delta^2-E^2}}{E_F}-\frac{n^2}{M^2}}$ are the
transverse wavenumbers of the electron in the S region and
$k_n(E)=k_F\sqrt{1+\frac{E}{E_F}-\frac{n^2}{M^2}}$ are the
transverse wavenumbers of the electron in the S region when
$\Delta=0$.  It is assumed that the Fermi wavenumber, $k_F =
\sqrt{2mE_F/\hbar^2}$ is the same in the S and N regions. All
the matrices are $M$ by $M$ dimensional. 
All the information on the normal region are incorporated in the matrix $G$
via the scattering matrix $S_0(E)$. 
Note that the secular equation (\ref{seceq:e}) is an extension of that 
derived for box and disk geometries in Ref.~\onlinecite{box_disk:cikk}. 

The density of states in the semiclassical Bohr-Sommerfeld 
approximation is written as~\cite{Gap-cikk}
\begin{eqnarray}
n(E) &=& M\int_0^{\infty}\!\mbox{d}s\,P(s)
\left[\frac{s}{\hbar v_F} + \frac{1}{\sqrt{\Delta^2-E^2}}\right]\nonumber\\
&&\times\sum_{n=0}^{\infty}\, \delta\left(\frac{s\,E}{\hbar v_F}-
\left(n\pi+\arccos\frac{E}{\Delta}\right) \right)
\label{DOS:eq}
\end{eqnarray}
This expression reduces to that of by Melsen et al.\ \cite{Melsen1}, 
Schomerus and Beenakker\cite{Andreev-Billiards}, 
Lodder and Nazarov\cite{Andreev-Billiards}, and 
Ihra et al.\ \cite{Richter1}, in the limit
$E \ll \Delta$ and when the coherence length in the superconductor, 
$\xi_0=\hbar v_{\rm F} /\Delta \ll d$. In that case, the  energy dependent 
phase shift $-\arccos(E/\Delta)$, due to Andreev reflections, 
was approximated by $\pi/2$. 
However, our expression is valid without such an approximation, and 
essentially improves the agreement between the exact and the 
Bohr-Sommerfeld approximation of the DOS 
(see also Ref.~\onlinecite{Gap-cikk}). 

From Eq.~(\ref{DOS:eq}) one can find a simple expression for 
the integrated DOS $N(E)= \int_0^E \, n(E^\prime)\, dE^\prime$: 
\begin{subequations}
\label{N-BS:e}
\begin{eqnarray}
N(E) &=& M \sum_{n=0}^\infty \, \bigl\{ 1-F\left[s_n(E)\right]\bigr\} , 
\,\, \text{where} 
\label{N-eq}  \\
s_n\left(E \right) &=& \frac{\left(n\pi + 
\arccos\frac{E}{\Delta}\right)}{E/\Delta}\,\, 
\xi_0 \; , 
\label{sn:eq}   \\
F(s) &=& \int_0^s \, P(s^\prime)\, ds^\prime,
\label{F-eq}
\end{eqnarray}
\end{subequations}
where $F(s)$ is the integrated distribution function of the return
probability of the electron. 
Note that $P(s)$ is normalized to one, i.e., $F(\infty)=1$. 

To obtain the exact energy levels of the AB shown 
in Fig.~\ref{geometria-fig},  
we need to calculate the scattering matrix $S_0(E)$ of the N region. 
Following the same approach as Mortensen and co-workers\cite{Mortensen2}, 
we obtain 
\begin{eqnarray}
S_0(E) &=& \varrho R(X,Q) \varrho^T -1, 
\,\,\,\,\, \text{where}  \\
R(X,Q) &=& 2{(1+Q)}^{-1}\Bigl\{1  \Bigr.  \nonumber \\
&&\Bigl. - 2 X^2 {\left[1+X^2+Q(1-X^2)\right]}^{-1} \Bigr\}. \nonumber 
\end{eqnarray}
Here $X$ is a diagonal matrix with elements 
$X_{mn}=\delta_{mn}\, \exp(ik_n d)$, 
$Q= \varrho^T \varrho$, and $\varrho$ is a 
$[k_{\rm F}W/\pi]$ by $[k_{\rm F}a/\pi]$ dimension matrix with 
elements given by the overlap integrals defined 
in Ref.\ \cite{Mortensen2} ($[.]$ stands for the integer part). 
Notice that if $W=a$ then $\varrho =1$ and 
the scattering matrix $S_0(E)=-X^2$ is a diagonal matrix. 
In this case, we obtain the same secular equation for AB
as that in Ref.~\onlinecite{box_disk:cikk} for box geometries. 

\begin{figure}[hbt]
\psfrag{qmlevels}{{\rm exact}}
\psfrag{BS}{{\rm BS}}
\psfrag{eps}[][][0.8]{$E/\Delta$}
\psfrag{N(E)}[][][0.8]{$N(E)$}
\includegraphics[scale=0.45]{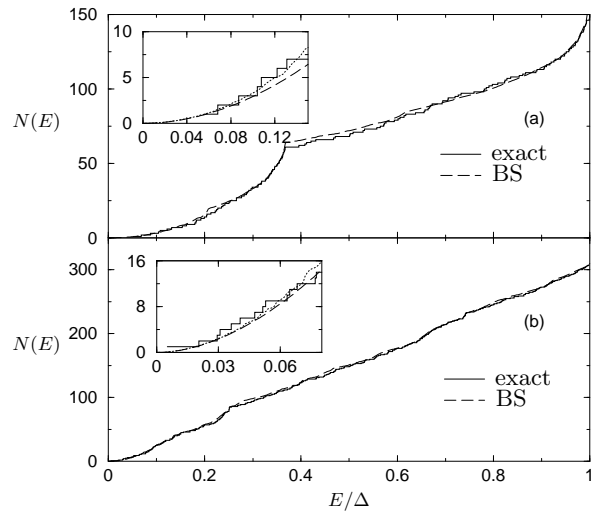}
\caption{The integrated density of states, $N(E)$ from exact quantum
mechanical calculations (solid lines) 
and from the Bohr-Sommerfeld approximation given by Eq.~(\ref{N-BS:e})
(dashed lines) as functions of $E$ (in units of $\Delta$) 
for $W=0.8a$ (a), and  $W=0.5a$ (b). Insets show the enlarged 
portions of $N(E)$ from exact calculations (solid line), 
its BS approximation  
from Eq.~(\ref{N-BS:e}) (dotted line) and its asymptotic form 
from  Eq.~(\ref{N_asympt:e}) (dashed line).
In both cases $d=a$ and the parameters are 
$M=55.5$, $\Delta/E_F = 0.015$. 
\label{Lepcsok-fig}}
\end{figure}

In Fig.~\ref{Lepcsok-fig} the exactly (numerically) computed 
integrated density of states obtained from Eq.~(\ref{exact-QM:e}) 
and its evaluation in Bohr-Sommerfeld (BS) approximation  
using Eq.~(\ref{N-BS:e}) are shown for different widths $W$ of the lead
(for the calculation of $P(s)$ see below). 
This shows that the exact calculations 
and the semiclassical predictions are in very good agreement.

\begin{figure}[hbt]
\psfrag{F}[][][0.8]{}
\psfrag{s}[][][0.8]{$s/d$}
\psfrag{x}[][][0.8]{$s/d$}
\psfrag{P(s)}[][][0.8]{$P(s)$}
\psfrag{F(s)BS}[][][0.8]{$\!\!\!\!\!\!F(s)$}
\psfrag{F(s)a}[][][0.8]{$F_a(s)$}
\psfrag{P(s)a}[][][0.8]{$P_a(s)$}
\psfrag{Y0}[][][0.8]{$0.0$}
\psfrag{Y1}[][][0.8]{$\;\;\;\;\;0.0005$}
\psfrag{0.0005}[][][0.8]{$\!\!0.0005$}
\psfrag{0.001}[][][0.8]{$0.001$}
\includegraphics[scale=0.4]{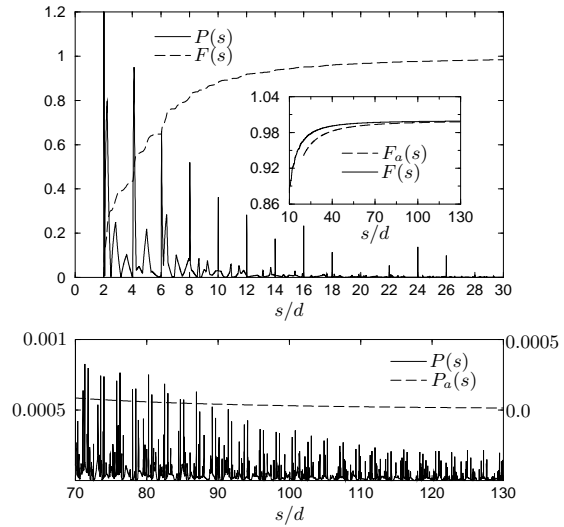}
\vspace*{-3mm}
\caption{Top panel: the return probability $P(s)$ (solid line) 
and its integral $F(s)$ (dashed line) as functions of $s/d$ 
for $W=a/2$ and $d=a$. 
In the inset $F(s)$ (solid line) and its asymptotic form $F_a(s)$ 
given by Eq.~(\ref{F_asymptotic}) (dashed line) are plotted 
as functions of $s/d$ for large $s$. 
Bottom panel: $P(s)$ (solid line and left $y$ axis) for larger $s$ 
along with the asymptotic expression of the return probability 
$P_a(s)=dF_a(s)/ds$ (dashed line and right $y$ axis). 
\label{PsFs-fig}}
\end{figure}

Calculations for $P(s)$ and $F(s)$ start with unfolding
the trajectory of the electron, i.e. using the fact that the free 
motion of the particle in the billiard is equivalent to its flight 
in a lattice of vertical intervals with length $W$ and with lattice
constant $2d$ ($a$) in the horizontal (vertical) direction.
In general, one can calculate $P(s)$ by using a large number of
trajectories and determining the distribution of their path lengths. 
An analytic form of $P(s)$ and $F(s)$ can be derived for $W\ge a/2$. 
As an example, these functions are plotted for $d=a$
and $W=a/2$ in Fig.~\ref{PsFs-fig}. 
Clearly, for arbitrary $W$, $P(s)=0$ for $s < 2d$ and 
for larger $s$ it is a
non-differentiable function possessing a singularity 
at $s=2d$ and peaks at  multiples of $2d$.
Using Eq.~(\ref{DOS:eq}) one can calculate the DOS and pronounced peaks 
(indeed singularities) arise due to the singularity of $P(s)$ 
at $s=2d$. The positions of these singularities in the DOS are 
in perfect agreement with that obtained from the general expression 
derived in Ref.~\onlinecite{box_disk:cikk}. 

The integrated path length distribution $F(s)$ is a smooth function of $s$ 
allowing one to obtain its asymptotic form $F_a(s)$ for $s \to \infty$ 
(see the inset of the top panel of Fig.~\ref{PsFs-fig}). 
From Eq.~(\ref{DOS:eq}) one can see that  
the large $s$ behavior of $P(s)$ is related to the low energy 
dependence of the DOS.
Due to the rapid variation of $P(s)$, it is difficult to define its asymptote. 
To avoid this problem we first calculate the asymptotic behavior of 
$F(s)$ and the asymptotic form of $P(s)$ is then calculated 
from $P_a(s)=dF_a(s)/ds$ (see also the bottom panel of Fig.~\ref{PsFs-fig}).  
For $s \to \infty$ analytical form of the 
asymptote of $F(s)$ is found by considering
how particles can travel a large distance $s$ without hitting the vertical
intervals of length $W$ in the unfolded space.
Such trajectories lie only in certain directions forming corridors 
with slopes $u$ that are multiples of $a/(2kd)$, where $k=[W/a]$.
The probability that a particle runs in such a corridor is
proportional to $1/(u s^2)$ for large $u$ but 
$|u| < u_{\rm max}=s/x_{\rm min}$, where 
$x_{\rm min}$ is a minimal displacement in the horizontal direction.  
Summation of the probabilities for these corridors yields a term
proportional to $s^{-2}\ln u_{\rm max}$.
Directions with slopes larger than $u_{\rm max}$ should be treated
separately and the result contains a term $s^{-2}\ln x_{\rm min}$.
The above terms from the two regions sum up to $s^{-2}\ln s$, and 
finally we find 
\begin{equation}
\label{F_asymptotic} 
\!\!\!\!F(s) \approx   F_a(s) \equiv 1
-\left(C_1 + C_2\, \ln \frac{s}{\sqrt{A}}\right) \frac{A}{s^2}
\end{equation}
for large $s$. Here $C_1$ and $C_2$ are functions of $d/a$ and $W/a$.
For the case $W \ge a/2$ the detailed calculation gives
analytic expressions also for $C_1$ and $C_2$:
\begin{subequations}
\begin{eqnarray}
C_1 &=& \!\frac{d}{a}\!\left(\!8\! -\! 6\frac{W}{a}\!\right)\! + C_2 \!  
\left[ c  - \frac{1}{2}\ln(d/a) \!\right]\! ,    \\
C_2 &=& \frac{4d}{W}\, {\left(1-\frac{W}{a}\right)}^2, 
\label{C2-a}  \\ 
c &=&\gamma + \frac{1}{2}+
2\sum_{j=1}^\infty \, K_0\left(4\pi j\, \frac{d}{a}\right).
\end{eqnarray}
\end{subequations}
Here $\gamma$ is the Euler-Mascheroni constant and $K_0$ is 
the modified Bessel function of the second kind. Therefore, 
$c$ is a weakly dependent function of $d$ for $d \ge a$. 

In the case of $W<a/2$ additional passages with slopes corresponding 
to noninteger multiples of $a/(2d)$ open and $C_1, C_2$ are modified.
$C_1$ is numerically determined, while for $C_2$ we have the formula
\begin{equation}
C_2 = \frac{4d}{W}\sum_{m=1}^{\left[a/W \right]} \frac{N_m}{m} \, 
\left(1-m \frac{W}{a} \right)^2\;,  
\label{C2-b}
\end{equation}
where $N_m$ is the number of those integers in the interval 
$[0,m-1]$ which are relative prime to $m$.
(Note that  $N_1=1$). 

Using the smooth asymptote of $F(s)$ given by Eq.~(\ref{F_asymptotic})
and Eqs.~(\ref{N-BS:e}), 
an analytic expression can be found for the low energy behavior of the
DOS: 
\begin{subequations}
\label{asym_N-rho}
\begin{eqnarray}
N_a(E) &=&  \frac{M}{8\pi^2} {\left(\frac{E}{E_T}\right)}^2  
\left[ \alpha +  \beta 
\left( \frac{1}{2}-\ln \frac{E}{E_T} \right)\right],  
\label{N_asympt:e}  \\
 n_a(E) \, \delta &=& \frac{1}{\pi} \, \frac{E}{E_T}\, 
\left( \alpha -  \beta \, \ln \frac{E}{E_T} \right),  
\label{asymDOS:e}  \\[1ex]
\alpha  &=& \frac{W^2}{A}\, \left[C_1  + C_2
\left(\kappa + \ln \frac{2\pi^2}{W/\sqrt{A}}\right) \right],  \\[1ex]
\beta  &=&  \frac{W^2}{A}\, C_2\, , 
\label{beta:def} \\[1ex]
\kappa &=&  -\frac{1}{2}-\frac{4}{3}\ln 2 -\frac{6}{\pi^2}\zeta^{\prime}(2) 
 \approx \! -0.854235,   
\end{eqnarray}
\end{subequations}
and $E_{\rm T}=M\delta/(4\pi)$ is the Thouless energy~\cite{Melsen1}. 
Besides the term proportional to the energy (which was also predicted in
Refs.~\onlinecite{Melsen1,Richter1}), there is  an additional 
logarithmic factor. Moreover, in contrast to the above mentioned
references the coefficient $\alpha$ depends on $W, a$ and $d$.
However, $\beta$ depends only on $W/a$.
Note that $W=a$ is a special case corresponding to the system studied by 
de~Gennes and Saint-James \cite{deGennes-Saint-James}. Then, $C_2=0$ and 
no logarithmic factor appears in the DOS and the result
is the same as in Ref.~\onlinecite{box_disk:cikk}.  

\begin{figure}[hbt]
\psfrag{Rodelta}{$n_a(E)\delta$}
\psfrag{ET}{$E/E_{\rm T}$}
\psfrag{alpha}{$\alpha$}
\psfrag{beta}{$\beta$}
\psfrag{Wa}{$W/a$}
\psfrag{r=1}{$d/a=1$}
\psfrag{r=1}{$d/a=2$}
\psfrag{r=1}{$d/a=1/2$}
\psfrag{r=1}{$d/a=1/4$}
\psfrag{r=1}{$d/a=1/8$}
\includegraphics[scale=0.5]{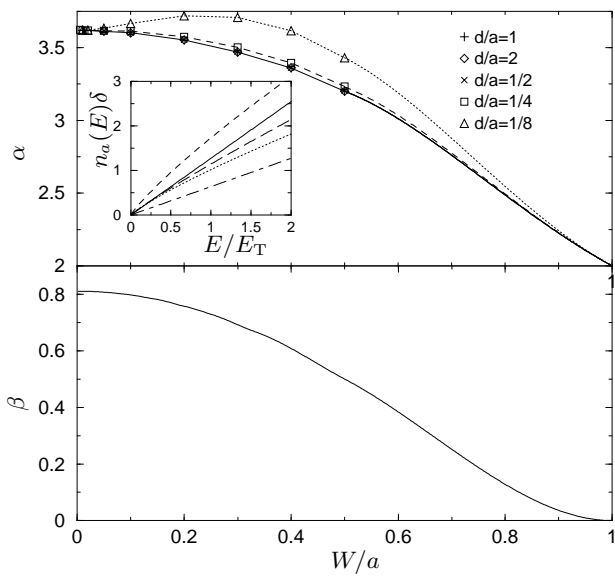}
\vspace*{-3mm}
\caption{Top panel: $\alpha$ as a function of $W/a$ 
for $d/a=1, 2, 1/2, 1/4, 1/8$. 
Bottom panel:  $\beta$ as a function of $W/a$ obtained from 
Eq.~(\ref{beta:def}) using Eqs.~(\ref{C2-a})
and (\ref{C2-b}) for $C_2$. 
The inset of the top panel shows the asymptotic DOS $n_a(E)$ 
(in units of $1/\delta$) for $d=a$, $W/a= 1, 0.5, 0.1, 0.01$ 
(dot-dashed, dotted, long-dashed, and dashed lines, respectively)
and the result from Ref.~\onlinecite{Melsen1}  
(solid line).}
\label{albe-fig}
\end{figure}

It is interesting to see the $W\ll a$ limit for which one would expect
that $P(s)$ has a {\em universal} limiting form, namely it is {\em only} 
a function of $W^2/A$. In this case, the electron
has enough time to explore the whole available phase space before
escaping (even in an integrable billiard) and $P(s)$ looses any detailed
dependence on the geometry of the billiard. Therefore, two billiards with 
the same $W^2/A$ but different aspect ratio ($d/a$) should have the same
path length distribution for small enough $W$. 
From numerical calculations we found that for $W \to 0$, $\alpha$ converges 
to a value $\alpha \approx 3.62$ independent of the aspect ratio, i.e., 
it becomes a {\em universal} constant. Similarly we find that 
when $W \to 0$, $\beta$
also tends to a {\em universal} constant $\beta \to 8/\pi^2$. 
These results are demonstrated in Fig.~\ref{albe-fig}.
In the inset of the top panel of Fig.~\ref{albe-fig} 
the asymptotic DOS $n_a(E)$ obtained from Eq.~(\ref{asymDOS:e}) is 
plotted for different values of $W$ along with the result 
from Ref.~\onlinecite{Melsen1} for the sake of comparison. 
Our results are in the same order of magnitude as that found in Melsen 
et al.~\cite{Melsen1} and Ihra et al.~\cite{Richter1}. 
However, from our analysis it turns out that the functional form of
the asymptotic DOS is {\em not} just a linear function but involves a
logarithmic factor. 

In conclusion, we have shown that exact quantum mechanical calculations for 
the integrated DOS of rectangular Andreev billiards agrees well 
(for the whole energy range below the gap) with 
that obtained from the Bohr-Sommerfeld approximation provided the energy 
dependent phase shift is taken into account. From the exact analytic form 
of the asymptote of the integrated return probability, we predict a new, 
logarithmic contribution to the DOS on the scale of the Thouless energy. 
In contrast to earlier results, we show that the DOS 
at this energy range explicitly depends on $W, a$ and $d$ but it has a  
universal limiting form for small enough $W$. 
We also investigated the case when the superconductor is placed anywhere 
at the side of the rectangle and found that the logarithmic
contribution in the DOS is generic for rectangular Andreev billiards. 

One of us (J.\ Cs.) gratefully acknowledges very helpful discussions
with C.~W.~J.~Beenakker. 
This work is supported in part by the EU's Human Potential
Programme under Contract No. HPRN-CT-2000-00144,
the Hungarian-British Intergovernmental Agreement on Cooperation in
Education, Culture, and Science and Technology, 
and the Hungarian  Science Foundation OTKA  TO34832.
One of us (Z. K.)  thanks the Hungarian Academy of Sciences 
for its support by a J\'anos Bolyai Scholarship.


\begin{thebibliography}{10}
\bibitem{note} In Refs.~\onlinecite{Melsen1,Richter1} the geometry of AB
can be obtained by cutting our AB at its symmetry axis.
Our results are irrelevant to this geometrical difference.
\end{thebibliography}


\end{document}